# Foreign-trained faculty and the collaborative organization of high-impact U.S. science

**Erjia Yan[1]*, Chaoqun Ni[2], Xiang Zheng[2], Weiye Gu[2]**

[1] Department of Information Science, College of Engineering and Computing, Drexel University, Philadelphia, PA 19104, USA

[2] Information School, University of Wisconsin-Madison, Madison, WI 53706, USA

*Corresponding author: Erjia Yan: ey86@drexel.edu

## Abstract

Internationally mobile scientists are central to national research and innovation systems. We link faculty rosters from the Academic Analytics Research Center to OpenAlex publication records for 2011–2020, yielding more than 12 million faculty–publication observations for 236,394 tenure-system faculty at more than 300 major U.S. universities. Foreign-trained faculty, defined by a terminal degree awarded outside the United States, constitute about 11% of the observed faculty workforce but account for 13–14% of publications and 14–16% of top-1% cited elite output. We find that this elevated representation in elite output is closely related to their organizational embeddedness: when faculty are compared within the same institution, scientific domain, rank, and year, the difference in elite output narrow substantially while overall productivity and collaboration differences persist within those settings. In addition, foreign-trained faculty enter each publication year with larger and broader prior collaboration networks, and collaborator reach is more strongly associated with subsequent elite output. By contrast, although raw topic breadth is greater among foreign-trained faculty, after accounting for prior publication volume, we find that their topic breadth is slightly narrower and more cognitively concentrated. These findings recast international training in the lens of scientific capacity and organizational integration in that internationally accumulated scientific capabilities become embedded in institutions and relationships through which research is organized and produced.

## 1. Introduction

National research systems increasingly utilize scientific capabilities formed beyond their borders (Akbaritabar et al., 2025; Rovito et al., 2021). While universities, research infrastructures, funding decisions, and pedagogical policies remain nationally organized, the scientists working within such structures can acquire training beyond the national boundaries (Gu et al., 2024; Han et al., 2024). Universities in the United States in particular employ faculty trained across a wide range of national research systems, who in turn contribute to various aspects of knowledge production, research team formation, and doctoral training (Elias et al., 2024; Merriweather et al., 2022; Wapman et al., 2022). International training is therefore not only a feature of scientific mobility or academic labor markets. It raises broader research policy question: how are scientific capabilities accumulated abroad incorporated into the institutions and collaborative structures of a national research system?

There is a rich set of literature that examines the association between foreign-born or foreign-trained status and research productivity and impact, finding that such status is often associated with higher productivity, greater international collaboration, and larger representation in highly cited research (Franzoni et al., 2012, 2014; Freeman, 2015; Jonkers & Cruz-Castro, 2013; Scellato et al., 2015; Verginer & Riccaboni, 2021). While these studies establish the importance of international mobility for knowledge productivity and impact, we have limited understanding of how those contributions are organized within U.S. institutions. Their elevated contributions may be the results of factors related to their organizational embeddedness, social structures of their collaboration networks, the diversity of their research portfolios, or combinations of these factors. To delineate such distinctions matters because scientific capacities are deeply intertwined in the environment they are made. Universities are highly stratified in resources, infrastructure, visibility, and access to scientific opportunities (Clauset et al., 2015; Morgan et al., 2018; Wapman et al., 2022; Way et al., 2019). Meanwhile, collaboration networks shape access to expertise, information, and research capacities (Jones et al., 2008; Wagner & Leydesdorff, 2005; Wuchty et al., 2007). For scientists trained in different national research systems, they may also possess different intellectual communities, research portfolios, and greater opportunities for knowledge recombination, compared with those trained domestically (Ganguli et al., 2019). Organizational placement and network position are therefore not simply covariates to be controlled away; they may constitute part of the process through which internationally accumulated capabilities become embedded in domestic scientific production.

We examine these possibilities by linking annual tenure-system faculty rosters from the Academic Analytics Research Center (AARC) to OpenAlex publication records from 2011 through 2020 (Priem et al., 2022). The resulting longitudinal panel contains 12,109,441 faculty–publication observations for 209,959 U.S.-trained and 26,435 foreign-trained faculty from more than 300 major U.S. universities. Training origin is defined by the location of the terminal-degree institution, and it does not imply nationality, citizenship, birthplace, or immigration status. We design three analytical levels. First, we examine the collective contributions of foreign-trained faculty by benchmarking their shares of publications and top-1% cited papers (i.e., elite output) against their share of the observed faculty workforce. Second, to understand the organizational factors associated with their elite output representations, we use decomposition and exact comparisons within institution × scientific domain × academic rank × year strata. Third, we gain understanding of the social and cognitive factors of their elite output representations by the reconstruction of collaboration networks and cognitive portfolios from the three years preceding each observational year and the examination of their relationships with subsequent elite output.

We find that foreign-trained faculty constitute approximately 11–12% of the observed faculty workforce but account for 13–14% of publications and 14–16% of elite output. This elevated representation in elite output is closely related to their organizational embeddedness: foreign-trained faculty are strongly represented in research-intensive institutions, and the difference in elite output narrow substantially when faculty are compared within the same institution, scientific domain, rank, and year, while the overall productivity and collaboration advantage persists within those settings. As for social factors, foreign-trained faculty enter each publication year with larger and broader prior collaboration networks, and collaborator reach is more strongly associated with subsequent elite output among foreign-trained faculty. By contrast, although raw topic breadth is greater among foreign-trained faculty, after accounting for prior publication volume, we find that their topic breadth is slightly narrower and more cognitively concentrated. Robustness and heterogeneity tests further reveal that these results are robust to alternative temporal windows and impact thresholds and is broadly similar across scientific domains, although the results vary across national training systems.

These findings recast international training in the lens of scientific capacity and organizational integration, while broaden the scope of academic mobility to the organization of national research systems. In addition, this study, through research designs that compare foreign-trained faculty with their comparable domestic counterparts, delineates within-context differences. Last, the comparison of social and cognitive organization identifies an important asymmetry: the more consistent distinction associated with foreign training lies in collaboration-network structure rather than unusually broad cognitive portfolios. The broader implication is that internationally accumulated scientific capabilities do not enter national research systems simply as additional units of human capital. They become embedded in institutions and relationships through which research is organized and produced.

## 2. Literature review and conceptual framework

### 2.1. International training, mobility, and scientific performance

International scientific mobility is seen as a main channel for human capital circulation and knowledge transfer between national research systems (Akbaritabar et al., 2025; Gu et al., 2024; Mihut et al., 2016). Early debates emphasized brain drain and brain gain, in which more advanced national research systems attract talents from less developed systems, thus causing a brain drain to the exporting systems while a brain gain to the importing systems (Bhagwati & Rodriguez, 1975; Stark et al., 1997; Weinberg, 2011). Subsequent work used the term brain circulation to better capture the complete trajectories of some migration scientists who received training and secured employment in more advanced foreign systems and later brought their expertise back to their home system (Jöns, 2009; Saxenian, 2005; Scellato et al., 2015). Studies in this venue also focused on the capacity of mobile scientists to maintain connections across origin and destination systems (Flanagan, 2015; Freeman, 2015). Internationally mobile scientists exhibit higher publication rates, greater citation impact, and more international collaboration than non-mobile researchers, although these relationships vary across fields, destinations, and stages of scientific careers (Franzoni et al., 2012, 2014; Gu et al., 2024; Han et al., 2024; Jonkers & Cruz-Castro, 2013; Scellato et al., 2015).

Research in science of science qualified this aggregate mobility premium. Jonkers and Cruz-Castro (2013) showed that international experience can persist in both collaboration patterns and research performance after scientists return to their home system. Scellato et al. (2015) demonstrated that migrant scientists maintain larger international networks, and the return on larger networks depends on the national research capacity of the home system. More recently, Benz and Larivière (2025) show that the mobility premium is strongly associated with combinations of geographic, institutional, academic, and disciplinary factors and there are marked diversity in the premium depending on which factors are involved.

International training constitutes a distinctive but comparatively understudied part of this process. Doctoral and other terminal-degree environments transmit not only knowledge and skills but also scientific norms, intellectual lineages, and professional relationships that may persist across subsequent careers (Capponi & Frenken, 2025). For foreign-trained faculty employed in the United States, these prior environments intersect with a highly stratified destination system. U.S. universities differ in research resources and scientific opportunities, and academic hiring and placement are an notable manifestation of this highly stratified system in that elite programs place large numbers of their graduates to other elite programs as well as programs at other tiers, while non-elite programs place graduates at other non-elite programs and often at a reduced tier (Clauset et al., 2015; Wapman et al., 2022). Consequently, differences in productivity and impact between foreign- and domestically trained scientists can reflect both prior training trajectories and their subsequent allocation across stratified organizational environments. Separating aggregate contribution from differences within comparable organizational settings therefore provides a way to connect the mobility literature with research on institutional stratification and cumulative advantage (Way et al., 2019; Zhang et al., 2022), hence the goal of this research.

## 2.2. Mobility, collaboration networks, and knowledge recombination

Another relevant body of literature identifies collaboration networks as a central channel through which mobility can influence scientific careers. As scientific production increasingly depends on teams and distributed expertise, mobile scientists are likely to possess this advantage by connecting knowledge, resources, visibility, and access to additional research communities at both the origin and destination move locations (Börner et al., 2010; Jones et al., 2008; Wuchty et al., 2007). International mobility therefore can expand these networks by allowing researchers to retain relationships from previous institutions and national research systems while developing new ties in their destination system.

Evidence reveals this collaboration mechanism and treats international connections as a form of social capital. Studies found that return scientists collaborate with researchers in their former host systems at an elevated rate (Jonkers & Cruz-Castro, 2013) while migrant scientists have larger international networks than native non-mobile researchers and they maintained persistent connections to scientists at home systems (Franzoni et al., 2014; Scellato et al., 2015). More recently, Chen et al. (2025) show that even indirect international connections, such as foreign ties gained through domestic collaborators, are associated with higher productivity and impact among junior researchers.

In addition to collaboration, access to diverse knowledge offers a different mechanism. Mobility may expose scientists to new intellectual traditions and knowledge bases, thus increasing opportunities for novel knowledge recombination (Franzoni et al., 2014; Gibson & McKenzie, 2014). Research on scientific novelty shows that atypical combinations of knowledge can generate work of higher impact (Uzzi et al., 2013). Similarly, interdisciplinary work that draw knowledge from diverse disciplinary sources with broader cognitive breadth can also boost impact (Larivière et al., 2015; Leahey et al., 2017). International training could therefore be associated with broader intellectual portfolios as well as broader social networks.

These mechanisms should not be conflated. International exposure may expand the range of people through whom knowledge is accessed without increasing the range of topics represented in their own research portfolio. Conversely, intellectual diversification may not require unusually extensive collaboration networks. Distinguishing social from cognitive breadth therefore provides leverage on a question left open by the mobility literature: whether the scientific contribution associated with international training is expressed primarily through broader individual knowledge portfolios or through the relational structure connecting scientists to distributed knowledge.

# 3. Data and methods

## 3.1. Data sources and record linkage

We constructed a longitudinal faculty–publication panel by linking annual faculty rosters from the Academic Analytics Research Center (AARC) to publication records from OpenAlex (Priem et al., 2022). AARC contains annual information on tenure-system faculty at more than 300 major U.S. universities, including person and institution identifiers, employing units, academic rank, disciplinary classification, and terminal-degree institution. We mapped the annual AARC snapshots to calendar years 2011–2020, which defines the main observation period for concurrent faculty employment and scientific output. The AARC disciplinary taxonomy contains 170 disciplines, which we classified into 28 broader areas and nine umbrella domains: Applied Sciences, Business and Economics, Education, Engineering, Humanities, Mathematics and Computing, Medicine and Health, Natural Sciences, and Social Sciences. AARC institutions were disambiguated and linked to the 2025 Carnegie Classification of Institutions of Higher Education to obtain institutional research intensity classification (American Council on Education, 2026).

To construct prior collaboration and cognitive portfolios, we used the OpenAlex works snapshot dated March 7, 2026 provided by SUB Göttingen, including work and author identifiers, publication year, authorship position, corresponding-author status, field-weighted citation impact (FWCI), field- and year-normalized top-10% and top-1% indicators, and the OpenAlex topic hierarchy. Publications whose year falls within 2011–2020 were aligned to the corresponding AARC roster year for analyses of scientific output. Publications preceding the observation year were used only when constructing prospective collaboration networks and cognitive portfolios. For instance, for the robustness analysis using a five-year collaboration window, the collaboration history is extended to 2006 so that the 2011 outcome year has a complete observational window.

For large science of science research, one key issue is linking faculty records (e.g., AARC) to publication authors (e.g., OpenAlex) while minimizing false author matches. We therefore used a one-to-one entity-resolution procedure that combines name, institutional, and disciplinary information. Exact normalized-surname blocking first reduced the candidate set to about on average eight OpenAlex authors to one AARC professor; candidate scores then incorporated surname and given-name agreement, first- and middle-initial consistency, institutional-affiliation similarity based on token overlap, and disciplinary overlap. Candidates were ranked within faculty member, and ambiguity was assessed using the separation between the highest- and second highest-scoring candidates. We retained high-confidence matches with a composite score $S \geq 0.80$ and a separation margin $\Delta \geq 0.05$, and medium-confidence matches with $0.50 \leq S < 0.80$ and $\Delta \geq 0.01$. Lower-confidence matches were excluded from analyses. Approximately 81% of AARC-covered faculty are linked at medium or high confidence. Data validity tests showed that shifting the matching thresholds by 0.02 changes the high-confidence set by less than 3%, although residual linkage error is possible, particularly for those with common names and sparse publication records. The resulting author–work data table contains 12,109,441 linked faculty–publication observations, covering 209,959 U.S.-trained and 26,435 foreign-trained faculty.

## 3.2. Training origin and outcome measures

We define faculty as foreign-trained when the terminal degree was awarded by an institution outside the United States and as U.S.-trained when the terminal degree was awarded by a U.S. institution. The determination was based on institution name recognition and web search. In the case an institution name alone was not enough to make the determination, we also utilized faculty information including their online C.V. and publication records. Fewer than 20 faculty records were ambiguous despite these efforts and were excluded from analyses. The degree records include 3,922 unique degree-granting institutions in 131 countries, including 1,321 U.S. institutions and 2,601 foreign institutions. Foreign-trained and U.S.-trained do not denote nationality, citizenship, birthplace, immigration status, ethnicity, or the complete trajectory of a scientist's education and career.

The primary measure is the OpenAlex top-1% cited indicator, referred to as elite output, defined using field- and year-normalized citation impact. At the faculty-year level, we summed the indicator across publications to obtain the number of top-1% papers produced by a faculty member in a given year. The top-10% indicator is used as a robustness outcome.

We also used several complementary outcomes to characterize scientific production and position, including publication output, mean field-weighted citation impact (FWCI), corresponding-authorship share, and mixed-training-origin team share that measures participation in collaboration that spans training origins.

### 3.3. System representation and organizational comparisons

### 3.3.1. System-level representation

For each year *t*, we calculate the share of observed faculty who were foreign-trained and compare this quantity with the foreign-trained share of publications, top-1% cited publications, and corresponding-authored publications. Because a publication can contain both foreign- and U.S.-trained faculty authors, we therefore use fractional allocation for system-level publication measures. A paper containing linked faculty from only one training-origin group is allocated entirely to that group, whereas a mixed-training-origin paper is divided equally between the two groups. To summarize representation at the upper tail of scientific impact, we define the annual elite representation ratio as

$$R_t^{elite} = \frac{S_{F,t}^{top1}}{S_{F,t}^{faculty}}$$

where $S_{F,t}^{top1}$ is the fractionally allocated foreign-trained share of top-1% cited publications in year *t*, and $S_{F,t}^{faculty}$ is the foreign-trained share of observed faculty in that year. A value of $R_t^{elite}$ above 1 indicate that their share of top-1% output exceeds their workforce presence. We construct analogous representation ratios for publications and repeat the calculations within scientific domains and institutional research-intensity categories.

### 3.3.2. Decomposition of the aggregate elite-output difference

We next examine how the observed difference in top-1% output changes when organizational factors are added sequentially. We estimate a series of nested logistic models in which top-1% status is the outcome and foreign-training origin is the focal predictor. Year fixed effects are included throughout, after which blocks representing scientific domain, institutional research intensity, institutional control, and career stage are introduced sequentially. Because coefficients from nested nonlinear probability models cannot be interpreted as a simple additive decomposition, we do not compare logit coefficients directly. Instead, for each specification *m*, we calculate a population-standardized foreign-trained-U.S.-trained contrast:

$$\Delta_m = \frac{1}{N}\sum_{n=1}^{N}\left[\hat{P}_m(Y_n = 1 \mid Foreign_n = 1, X_n) - \hat{P}_m(Y_n = 1 \mid Foreign_n = 0, X_n)\right]$$

where $Y_n$ denotes top-1% status, $Foreign_n$ denotes foreign-training status, and $X_n$ contains the covariates included in specification *m*. For every observation, we predict the probability of top-1% status once under foreign-trained status and once under U.S.-trained status while retaining the observation's actual values of all other covariates. Averaging these counterfactual predictions over the observed covariate distribution places all nested models on the same probability scale.

### 3.3.3. Within-context organizational comparisons

Because faculty may occupy very different combinations of institutions, fields, ranks, and historical periods, we construct strata defined by the Cartesian product of employing institution, scientific domain, academic rank, and calendar year:

$$s = Institution \times Domain \times Rank \times Year$$

A stratum is retained only if it contains at least one foreign-trained and one U.S.-trained faculty member. This procedure yields 38,223 eligible strata and ensures that every comparison is made between faculty working at the same institution, in the same broad scientific domain, at the same academic rank, and in the same year.

For outcome Y, the foreign-trained-U.S.-trained contrast within stratum *s* is

$$\Delta_s^Y = \overline{Y}_{F,s} - \overline{Y}_{U,s}$$

The within-context estimate is the equal-stratum average,

$$\Delta_{within}^Y = \frac{1}{S}\sum_{s=1}^{S} \Delta_s^Y$$

where S is the number of eligible strata. We apply this procedure to annual publication output, top-1% output, mean FWCI, participation in mixed-training-origin teams, degree centrality, and corresponding-authorship share.

## 3.4. Collaboration-network measures and models

We next examine whether foreign-trained and U.S.-trained faculty occupy different positions in the collaboration structure of U.S. science and whether prior network position is differently associated with subsequent elite output. The research design separates network measurement temporally from the outcome: collaboration networks are constructed exclusively from publications preceding the focal outcome year.

### 3.4.1. Construction of prior collaboration networks

For faculty member *i* observed in focal year *t*, we identify all coauthors appearing on that faculty member's OpenAlex publications during the preceding three years, *t*-3 through *t*-1. The focal faculty member is excluded from the collaborator set. Let $J_{it}$ denote the set of distinct collaborators observed in this prior window and let $w_{ijt}$ denote the number of publications coauthored by faculty member *i* and collaborator *j* during the window.

We characterize each prior network using three complementary measures: reach, breadth, and tie depth.

Collaborator reach is the number of distinct prior collaborators:

$$Reach_{it} = |J_{it}| = \sum_j \mathbf{1}\left(w_{ijt} > 0\right)$$

Collaborator breadth measures the effective number of collaborators after accounting for the concentration of collaboration across ties:

$$Breadth_{it} = \frac{\left(\sum_j w_{ijt}\right)^2}{\sum_j w_{ijt}^2}$$

Reach and breadth capture different aspects of network structure: reach counts accessible collaborators, whereas breadth discounts a nominally large network when activity is concentrated in a limited set of ties.

Tie depth captures repeated collaboration and is defined as

$$Depth_{it} = \frac{\sum_j w_{ijt}}{Reach_{it}}$$

Thus, tie depth is the mean number of shared publications per distinct collaborator. These measures distinguish having more collaborators, distributing collaborative activity more broadly across those collaborators, and repeatedly working with the same collaborators.

### 3.4.2. Training-origin differences in prior network position

For the network-position analysis in Fig. 3A, we retain faculty-years with a nonempty prior collaboration network. This produces 1,305,841 faculty-years representing 213,119 faculty.

Because reach and breadth are strongly right-skewed, they are log-transformed before standardization. For reach,

$$Reach_{it}^{*} = \frac{\ln(1 + Reach_{it}) - \mu_R}{\sigma_R}$$

and breadth is transformed in a similar manner. Tie depth is standardized directly. For each network dimension $Z_{it}$, we estimate

$$Z_{it} = \alpha + \beta Foreign_i + \gamma_t + \delta_{d(i,t)} + \rho_{r(i,t)} + \varepsilon_{it}$$

where $Foreign_i$ indicates foreign terminal-degree training, $\gamma_t$ is year fixed effects, $\delta_{d(i,t)}$ is scientific-domain fixed effects, and $\rho_{r(i,t)}$ is institutional research-intensity fixed effects. Because the dependent variables are standardized, beta represents the adjusted difference between foreign-trained and U.S.-trained in standard-deviation units. Standard errors are clustered at the faculty level to account for repeated observations of the same faculty member.

### 3.4.3. Prior network reach and subsequent elite output

The second network analysis asks whether prior collaborator reach is associated with subsequent elite scientific output and whether this association differs according to training origin. The analytical unit is the faculty-year. To enter this analysis, faculty member *i* must have a nonempty prior collaboration network and at least one publication in focal year *t*. The resulting sample contains 1,119,314 faculty-years from 200,348 faculty and 7,630,784 current-year publications.

Let $E_{it}$ denote the number of faculty member *i*'s publications in year *t* classified in the top 1% elite output, and let $N_{it}$ denote that faculty member's total number of publications in year *t*. We model

$$E_{it} \sim \text{Binomial}(N_{it}, p_{it})$$

where $p_{it}$ is the modeled probability that a current-year publication is a top-1% publication. The primary specification is

$$\text{logit}(p_{it}) = \alpha + \beta_1 Foreign_i + \beta_2 R_{it} + \beta_3 (Foreign_i \times R_{it}) + \gamma_t + \delta_{d(i,t)} + \rho_{r(i,t)}$$

The coefficient $\beta_2$ represents the association between a one-standard-deviation increase in prior network reach $R_{it}$ and the odds of elite output among U.S.-trained faculty. For foreign-trained faculty, the corresponding log-odds association is $\beta_2 + \beta_3$. Thus,

$$OR_{reach,U.S.} = e^{\beta_2}$$

$$OR_{reach,foreign} = e^{\beta_2 + \beta_3}$$

$$OR_{interaction} = e^{\beta_3}$$

An interaction odds ratio (OR) greater than 1 indicates that the association between prior network reach and subsequent top-1% output is stronger among foreign-trained than among U.S.-trained faculty. Inference uses an author-clustered covariance matrix constructed from observation-level score contributions aggregated within faculty, accounting for repeated faculty-year observations. We additionally report population-standardized predicted probabilities to supplement the OR reporting. For selected values *r* of standardized reach and training-origin status *f*, we calculate

$$\hat{P}(f, r) = \frac{1}{N} \sum_{n=1}^{N} \hat{P}\,(E = 1 \mid Foreign = f, R = r, X_n)$$

where every observed faculty-year is assigned the specified training-origin and reach values while retaining its observed year, scientific domain, and institutional research intensity. Predictions are then averaged across the analytical population. This marginal-standardization procedure preserves the observed distribution of organizational covariates while varying only training origin and network reach. Confidence intervals are obtained by parametric simulation from the author-clustered coefficient covariance matrix.

## 3.5. Cognitive-portfolio measures and models

We next examine its cognitive organization by asking whether foreign- and U.S.-trained faculty differ in the breadth and intellectual dispersion of their prior research portfolios and how these dimensions are associated with subsequent elite output.

As with the collaboration analysis, cognitive measures are constructed prospectively. For each faculty member *i* observed in focal year *t*, the cognitive portfolio consists of publications produced during *t*-3 through *t*-1, whereas elite output is measured in year *t*. Each prior publication is assigned its OpenAlex primary topic. The cognitive analysis therefore separates the structure of prior research from the subsequent outcome.

### 3.5.1. Topic breadth

Let $n_{ikt}$ denote the number of faculty member $i$'s prior-window publications assigned to topic *k*, and let

$$N_{it}^{prior} = \sum_{k} n_{ikt}$$

denote the total number of publications in the prior portfolio. The share of the portfolio represented by topic *k* is

$$p_{ikt} = \frac{n_{ikt}}{N_{it}^{prior}}$$

We measure raw topic breadth using exponential Shannon diversity:

$$B_{it} = \exp\left(-\sum_{k} p_{ikt} \ln p_{ikt}\right)$$

A portfolio concentrated entirely in one topic has $B_{it}$ = 1; a portfolio distributed equally across K topics has $B_{it}$ = K.

A central measurement issue is that observed topic breadth mechanically increases with the number of publications available from which topics can be observed. In the data, prior portfolio size explains a substantial portion of variation in raw breadth (R-squared = 0.49). We therefore treat portfolio-size adjustment as part of measurement construction rather than as a current-year productivity control.

Specifically, we estimate a flexible relationship between log topic breadth and prior publication volume using a cubic polynomial in log portfolio size and year fixed effects:

$$\ln B_{it} = \alpha + \theta_1 \ln N_{it}^{prior} + \theta_2 \left(\ln N_{it}^{prior}\right)^2 + \theta_3 \left(\ln N_{it}^{prior}\right)^3 + \gamma_t + u_{it}$$

The residual

$$\tilde{B}_{it} = \ln B_{it} - \widehat{\ln B}_{it}$$

measures whether an author's observed cognitive breadth is greater or smaller than expected for a prior publication portfolio of that size in that year. We standardize this residual:

$$B_{it}^{*} = \frac{\tilde{B}_{it} - \mu_{\tilde{B}}}{\sigma_{\tilde{B}}}$$

Positive values indicate a broader-than-expected portfolio given prior publication volume, whereas negative values indicate a narrower-than-expected portfolio.

### 3.5.2. Cognitive span

For every pair of distinct topics *k* and *l* represented in an author's prior portfolio, let $d_{kl}$ denote their hierarchical distance. A distance of 1 means different topics in the same subfield, 2 is different subfields in the same field, 3 is different fields in the same domain, and 4 is different domains. Cognitive span is the topic-share-weighted mean distance across distinct topic pairs:

$$Span_{it} = \frac{\sum_{k<l} p_{ikt}\, p_{ilt} d_{kl}}{\sum_{k<l} p_{ikt}\, p_{ilt}}$$

Thus, two portfolios can have comparable effective topic breadth while differing substantially in how far apart their constituent topics are in the scientific knowledge hierarchy. The weighting gives greater influence to topic pairs representing larger shares of the author's research portfolio. Cognitive span is standardized over the analytical sample.

### 3.5.3. Training-origin differences in cognitive portfolios

We first estimate whether adjusted topic breadth and cognitive span differ according to training origin. For cognitive dimension $C_{it}$, we estimate

$$C_{it} = \alpha + \beta Foreign_i + \gamma_t + \delta_{d(i,t)} + \rho_{r(i,t)} + \varepsilon_{it}$$

Because both outcomes are standardized, beta represents the adjusted foreign-trained-U.S.-trained difference in standard-deviation units. Models adjust for year, scientific domain, and institutional research intensity, with standard errors clustered by faculty member.

### 3.5.4. Joint cognitive configuration and elite output

We then examine whether breadth and cognitive span jointly characterize subsequent elite output. For faculty-years with at least one current-year publication, we again model the number of top-1% papers as grouped-binomial data:

$$E_{it} \sim \text{Binomial}(N_{it}, p_{it})$$

The cognitive model is

$$\text{logit}(p_{it}) = \alpha + \beta_1 B^*_{it} + \beta_2 Span^*_{it} + \beta_3 (B^*_{it} \times Span^*_{it}) + \beta_4 Foreign_i + \gamma_t + \delta_{d(i,t)} + \rho_{r(i,t)}$$

Here, $\beta_3$ tests whether the association between cognitive breadth and elite output depends on how intellectually dispersed the constituent topics are. Training origin enters as an adjustment variable rather than as part of the focal cognitive interaction. Accordingly, this model characterizes the relationship between prior cognitive configuration and elite output while accounting for training origin and organizational context. Confidence intervals are calculated based on 1,000 draws from the fitted coefficient distribution using the author-clustered covariance matrix.

## 3.6. Robustness, heterogeneity, and interpretation

### 3.6.1. Robustness to network and outcome definitions

the robustness tests, in addition to the main analysis of the three-year window, we reconstruct the exposure using one-, three-, and five-year lookback windows to test sensitivity to this choice. Collaboration histories extend to 2006, providing a complete five-year history for the earliest outcome year. We estimate each window using both its full eligible sample and a common sample of 980,186 faculty-years observed under all three definitions. We separately test whether the result depends on defining elite output at the extreme upper tail of citation impact. The primary top-1% outcome is replaced with the top-10% indicator while holding the 1,119,314 faculty-year sample, network exposure, fixed effects, and clustered inference constant.

### 3.6.2. Heterogeneity across scientific and organizational contexts

We examine whether the network interaction varies across career stage, institutional environment, scientific domain, and foreign training system. Career-stage models are estimated separately for assistant, associate, and full professors. Institutional analyses distinguish the highest-research institutions from other classified institutions. Scientific-domain heterogeneity is examined across the nine umbrella domains. In addition to domain-specific estimates, a pooled model includes training origin × network reach × domain interactions, and an eight-degree-of-freedom Wald test evaluates whether the foreign-trained × reach interaction differs jointly across domains. Last, we examine the 12 foreign terminal-degree countries represented by at least 500 faculty in the primary network sample: the United Kingdom, Canada, Germany, China, India, Russia, France, Japan, Italy, Australia, Israel, and Switzerland. Country-specific estimates compare each training system with the common U.S.-trained reference group. A pooled model provides the joint covariance matrix, from which an 11-degree-of-freedom Wald test evaluates equality of the 12 country-specific interaction coefficients; a secondary test evaluates whether they are jointly zero.

## 4. Results

### 4.1. Contribution relative to workforce presence

We first provide a high-level understanding of the contributions of foreign-trained faculty. Results show that across 2011–2020, they constitute approximately 10–11% of tenure-system faculty in the linked AARC panel while accounting for 13–14% of fractionally allocated publications (Fig. 1A). Their representation rises further in the upper tail of citation impact: foreign-trained faculty account for 14–16% of top-1% cited papers in every year (Fig. 1B). The elite representation ratio, which is their share of top-1% output divided by their workforce share, remains above 1 throughout the decade and increases from 1.34 in 2011 to 1.48 in 2020 (Fig. 1C).

The pattern is also visible in scientific leadership. Foreign-trained faculty account for a larger share of corresponding authors than of the faculty workforce and an even larger share of corresponding authors on top-1% papers (Fig. 1D). These comparisons show that the contribution is consistent throughout the observed window and that it is visible in several aspects including total production, elite output, and corresponding authorship.

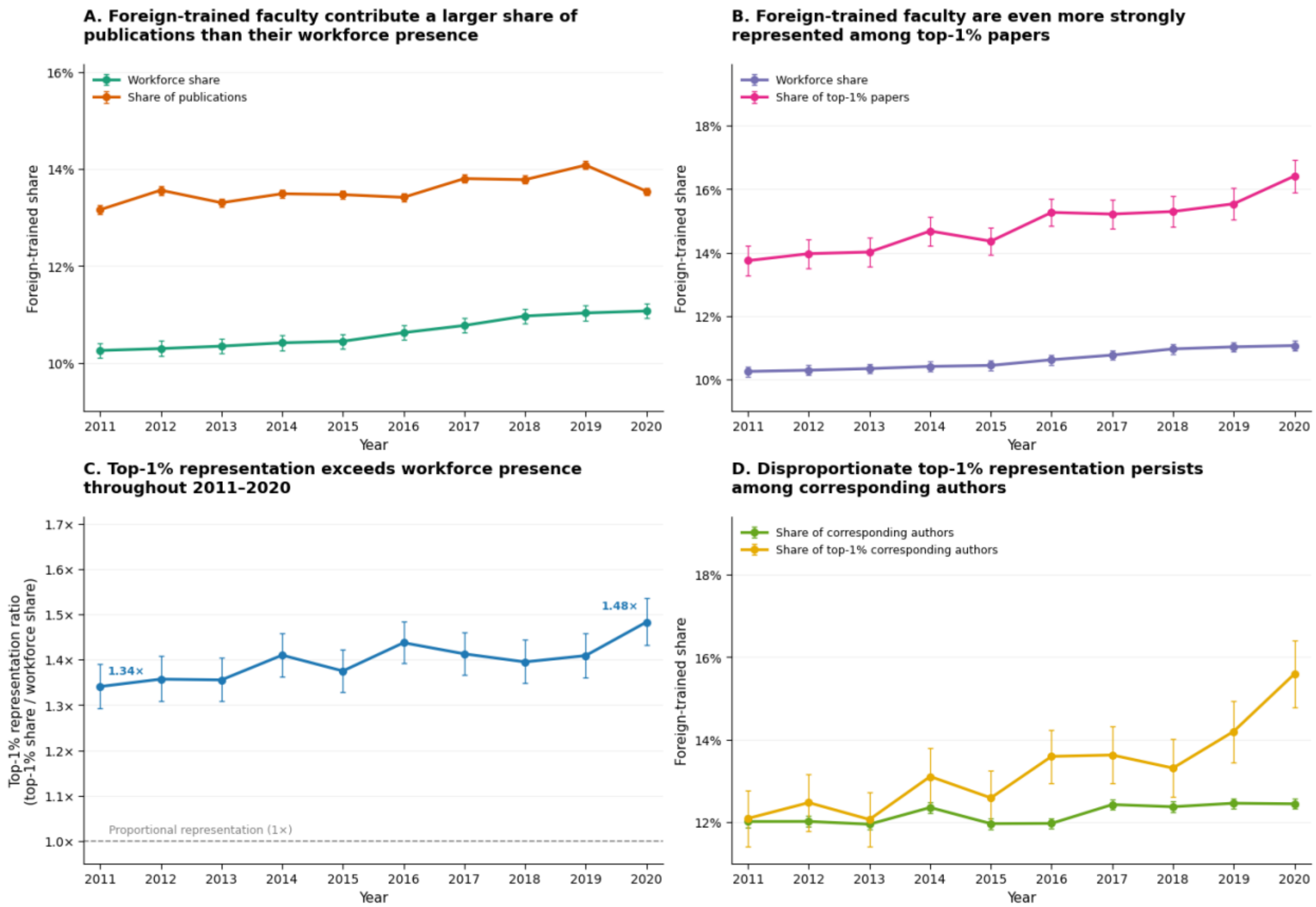

**Fig. 1.** Foreign-trained faculty contribute beyond their workforce presence, 2011–2020. (A) Annual workforce share and fractionally allocated publication share. (B) Workforce share and fractionally allocated share of top-1% cited papers. (C) Elite representation ratio, defined as the foreign-trained share of top-1% papers divided by the foreign-trained faculty share; 1 denotes proportional representation. (D) Foreign-trained shares among corresponding authors and corresponding authors of top-1% papers. Mixed-training-origin papers contribute one-half to each group in system-level shares; error bars show approximate 95% confidence intervals.

## 4.2. Organizational position and within-context differences

We next examine foreign-trained faculty's elite output share across different domains and institutional settings. Foreign-trained publication shares exceed workforce shares in most scientific domains, and top-1% representation is especially high in several STEM and social-science domains, with Mathematics and Computing and Humanities being the two notable exceptions (Fig. 2A). In terms of institutional settings, representation is also strongest in the highest- and high-research-intensity institutions and is visible in both public and private nonprofit universities (Fig. 2B). Foreign-trained faculty are therefore disproportionately present in parts of the system that produce large volumes of influential research.

We further investigate foreign-trained faculty's elevated representation in elite output using two methods, one is the nested controls, and the other is the within stratum comparison. Through nested controls, we find that sequentially adding domain, research intensity, institutional control, career stage, and year associates 13.2% of the differences in elite output, while leaving 86.8% of the differences outside the measured coarse composition (Fig. 2C; Table 1, Panel A). Within the identified sources of differences, institutional control contributes the largest positive component, while research intensity and year partially offset the observed gap.

We then use within strata to create stricter comparisons. Within 38,223 strata defined by the same institution, scientific domain, academic rank, and year, differences in elite output per faculty-year, mean FWCI, and corresponding-author share are greatly reduced (Fig. 2D; Table 1, Panel B). Combining the results from both nested decomposition and within stratum comparisons, we find that fine-grained exact institution match helps explain much of the differences, compared with the results from the decomposition that uses coarse institution categories.

Other outcomes persist within strata. Foreign-trained faculty publish 1.13 more papers per faculty-year in the raw comparison and 0.89 more within strata, retaining nearly four-fifths of the original difference. Their mixed-training-origin team share is 0.30 higher in the raw comparison and 0.27 higher within strata. These results separate two dimensions of the aggregate contribution: citation impact including elite output and mean FWCI are tightly linked to organizational location, while publication volume and cross-training-origin collaboration remain distinct among faculty operating in closely comparable settings.

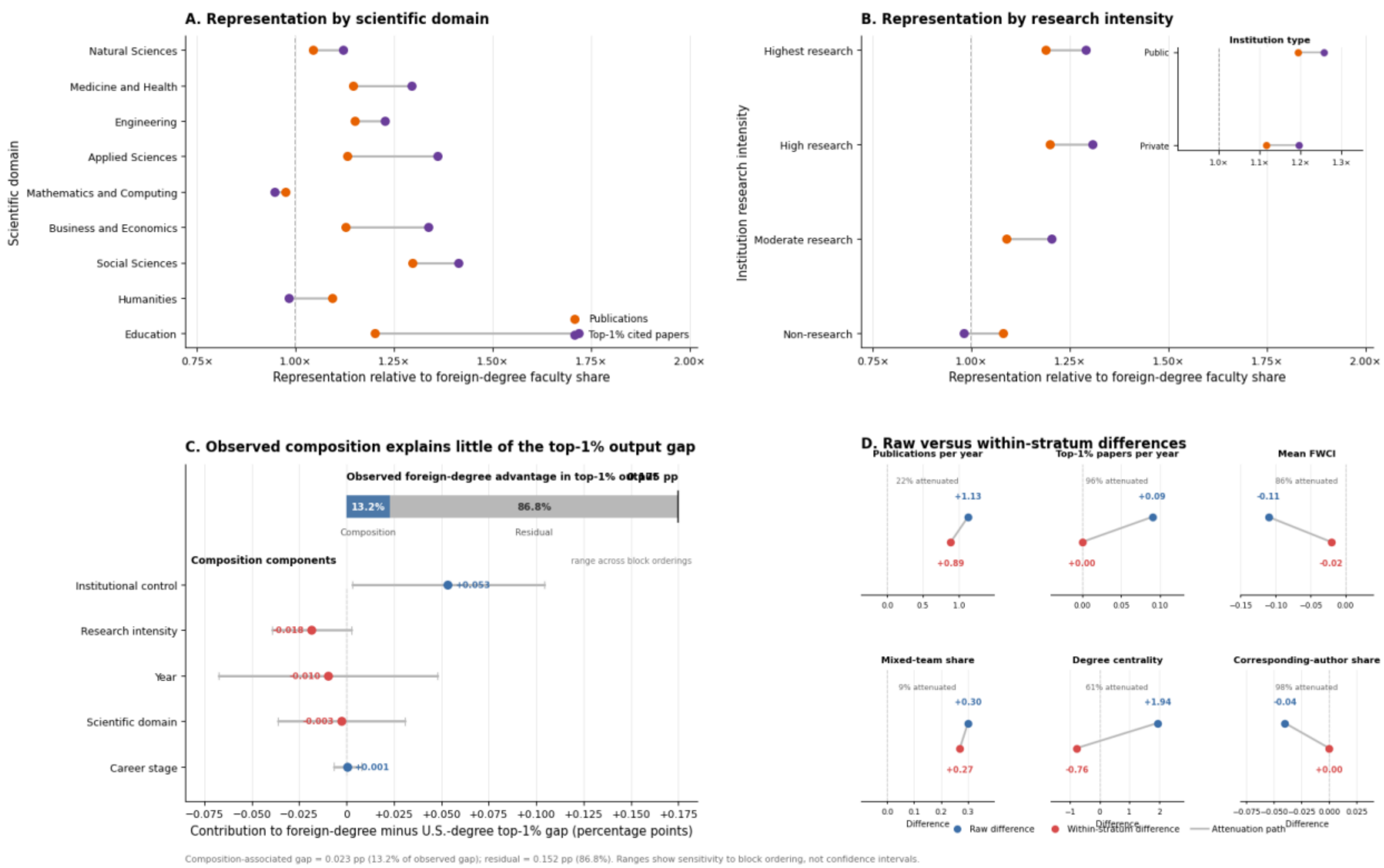

**Fig. 2.** Scientific impact, productivity, and organizational position. (A) Publication and top-1% representation ratios by scientific domain; values above 1 indicate output shares exceeding workforce shares. (B) Representation ratios by institutional research intensity, with an inset by public and private nonprofit control. (C) Sequential decomposition of the observed foreign-trained minus U.S.-trained top-1% gap. (D) Raw and within-stratum differences in publications, top-1% papers, mean field-weighted citation impact, mixed-training-origin team share, degree centrality, and corresponding-author share. Within-stratum contrasts compare faculty in the same institution, domain, rank, and year.

**Table 1.** Organizational decomposition and within-context comparisons

| Outcome or quantity | Observed / raw | Composition / within-stratum | Change |
|---|---|---|---|
| **Panel A. Sequential decomposition of the paper-level top-1% gap** | | | |
| Foreign-trained minus U.S.-trained gap | +0.175 pp | — | 100% |
| Composition-associated total | — | +0.023 pp | 13.2% |
| Residual | — | +0.152 pp | 86.8% |

| Outcome or quantity | Observed / raw | Composition / within-stratum | Change |
|---|---|---|---|
| **Panel B. Exact institution × domain × rank × year comparisons (38,223 strata)** | | | |
| Publications per faculty-year | +1.13 | +0.89 | 22% attenuated |
| Top-1% papers per faculty-year | +0.09 | ≈0.00 | 96% attenuated |
| Mean field-weighted citation impact | −0.11 | −0.02 | 86% attenuated |
| Mixed-training-origin team share | +0.30 | +0.27 | 9% attenuated |
| Degree centrality | +1.94 | −0.76 | Direction reverses |
| Corresponding-author share | −0.04 | ≈0.00 | 98% attenuated |

*Notes.* Differences are foreign-trained minus U.S.-trained. Panel A uses population-standardized probabilities from nested logistic models; component shares are descriptive and ordering sensitive. Panel B gives equal-stratum means among strata containing both training-origin groups. The two panels use different units and estimands and are not causal decompositions.

### 4.3. Prior collaboration networks and elite output

In this section, we report results on the association between faculty's prior collaboration network patterns and subsequent elite output. Results show that foreign-trained faculty enter the focal publication year with more extensive prior collaboration networks. Over the preceding three years, they have an average of 89.1 distinct collaborators compared with 64.8 among U.S.-trained faculty. They also have wider collaborator breadth at 54.0 compared with 40.0 for U.S.-trained faculty, and slightly deeper tie depth at 1.74 versus 1.66 shared papers per collaborator (Fig. 3A).

Adjustment for year, domain, and institutional research intensity reduces but does not remove these differences. Relative to U.S.-trained faculty, foreign-trained faculty have 0.130 standard deviations greater log collaborator reach (95% CI, 0.118–0.143), 0.104 standard deviations greater log collaborator breadth (95% CI, 0.092–0.116), and 0.037 standard deviations greater tie depth (95% CI, 0.029–0.046) (Table 2, Panel A). Network position is therefore a persistent dimension of difference even after broad scientific and organizational composition is considered.

Prior network reach is positively associated with subsequent elite output for both groups. A one-standard-deviation increase in standardized log reach is associated with an odds ratio of 1.359 among U.S.-trained faculty and 1.483 among foreign-trained faculty (Table 2, Panel B). The foreign-trained × reach interaction is 1.091 (95% CI, 1.054–1.130; $p < 0.001$), indicating that the reach association is approximately 9% larger on the odds-ratio scale for foreign-trained faculty.

Population-standardized probabilities clarify the magnitude and crossover pattern (Fig. 3B). At low prior reach (−1 SD), the adjusted probability of top-1% output is 0.343 percentage points lower for foreign-trained faculty. At average reach, the gap narrows to −0.153 percentage points. At high reach (+1 SD), it reverses to a 0.230-percentage-point foreign-trained advantage. The result shows that the collaboration premium is the most pronounced for foreign-trained faculty as their network reach is elevated to the highest levels.

Because the exposure precedes the outcome, current-year publications do not generate the measured network. The model nevertheless remains associational. Productive scientists may accumulate both larger networks and later high-impact work, and unobserved funding or team roles may influence both.

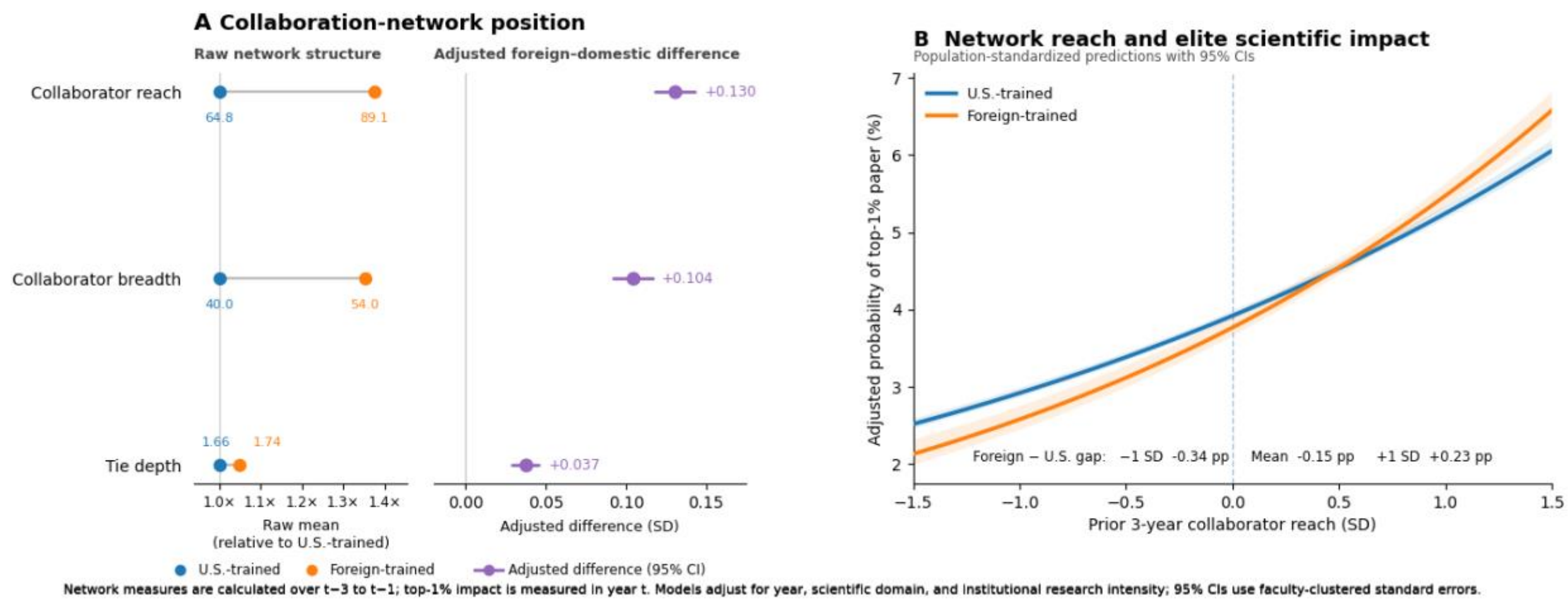

**Fig. 3.** Collaboration-network position and high-impact scientific output. (A) Raw prior-three-year collaborator reach, breadth, and tie depth by training origin, together with adjusted foreign-trained minus U.S.-trained differences and 95% confidence intervals. (B) Population-standardized top-1% probability by standardized log prior-three-year network reach. The foreign-trained × reach interaction is OR 1.091 (95% CI, 1.054–1.130); the adjusted origin gap changes from −0.34 percentage points at −1 SD reach to +0.23 percentage points at +1 SD.

## 4.4. Cognitive portfolios provide a contrasting pattern

In this section, we examine if there is an association between faculty's cognitive portfolios and their elite output. The key finding is that the cognitive results do not mirror the network results once publication volumes are controlled for. As measured by raw number of topics, foreign-trained faculty work across 7.48 topics in the prior three-year window, compared with 7.14 for U.S.-trained faculty (Fig. 4A). Raw breadth, however, is strongly related to the number of prior publications available for topic assignment. Once breadth is residualized on prior portfolio size and year, foreign-trained faculty have slightly narrower portfolios (−0.0487 SD; 95% CI, −0.0648 to −0.0326). Their cognitive span is also modestly lower (−0.0426 SD; 95% CI, −0.0564 to −0.0288), indicating slightly more coherent topic portfolios after broad field and institutional composition is adjusted (Table 2, Panel C).

As for the association with elite output we find that elite output is highest for narrow, coherent portfolios; broadening the portfolio is associated with lower top-1% probability, but this negative association diminishes as the topics in the portfolio become more cognitively distant (Fig. 4B). Adjusted breadth and cognitive span are each negatively associated with top-1% probability (OR = 0.946, 95% CI: 0.930–0.962; OR = 0.940, 95% CI: 0.926–0.954, respectively), while their positive interaction (OR = 1.037, 95% CI: 1.018–1.056) indicates that the negative association with breadth weakens as cognitive span increases.

The results suggest that foreign-trained faculty are more extensively connected across collaborators, and network reach is more strongly associated with elite output for them, but their topic portfolios are not broader than those of U.S.-trained faculty once the number of observable topics is equalized. The stronger distinguishing signal is social rather than cognitive breadth.

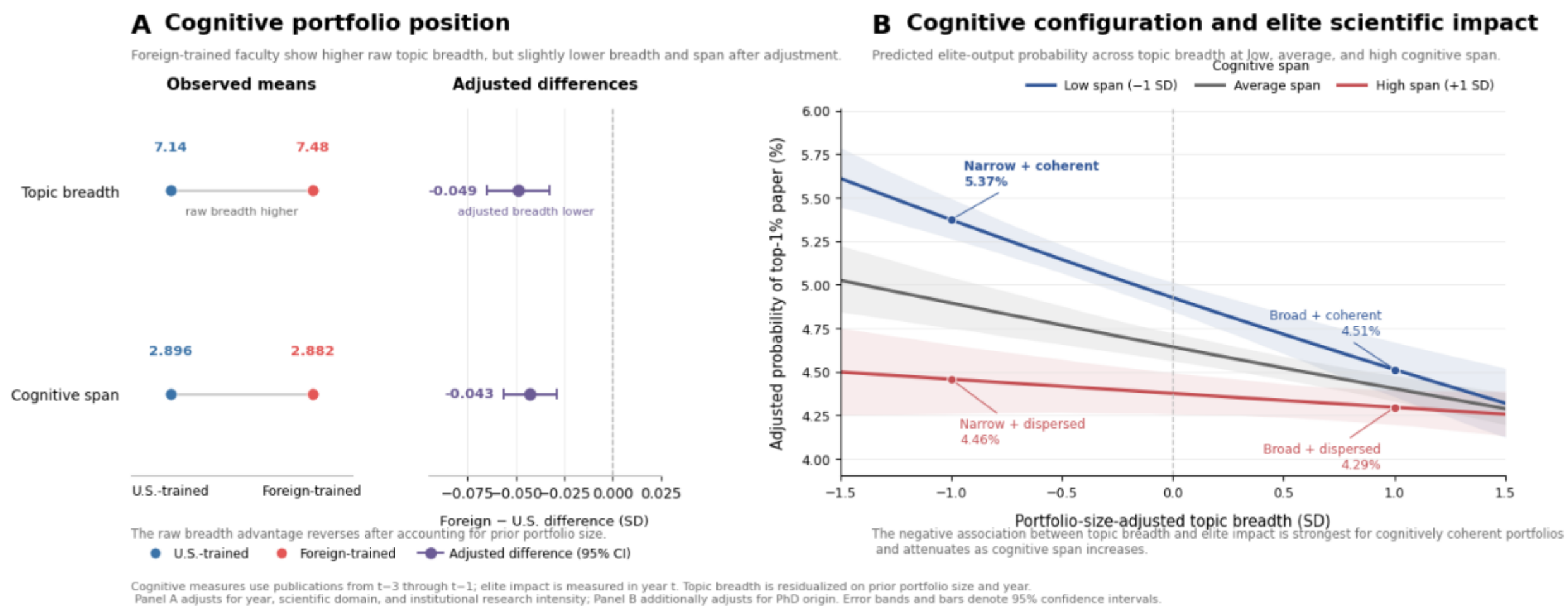

**Fig. 4.** Cognitive portfolio structure and high-impact scientific output. (A) Raw topic breadth and cognitive span by training origin and adjusted foreign-trained minus U.S.-trained differences with 95% confidence intervals. Adjusted breadth is residualized on prior portfolio size and year. (B) Population-standardized top-1% probability across adjusted breadth at low, average, and high cognitive span. The positive breadth × span interaction (OR 1.037; 95% CI, 1.018–1.056) indicates that the negative breadth association weakens as span increases.

**Table 2.** Main collaboration-network and cognitive-portfolio estimates

| Term | Estimate | 95% confidence interval | p | Scale / role |
|---|---|---|---|---|
| **Panel A. Adjusted foreign-trained minus U.S.-trained network-position differences** | | | | |
| Log collaborator reach | +0.130 | 0.118 to 0.143 | — | SD difference |
| Log collaborator breadth | +0.104 | 0.092 to 0.116 | — | SD difference |
| Tie depth | +0.037 | 0.029 to 0.046 | — | SD difference |
| **Panel B. Prior network reach and current-year top-1% output** | | | | |
| Reach association, U.S.-trained faculty | 1.359 | — | — | Odds ratio |
| Reach association, foreign-trained faculty | 1.483 | — | — | Odds ratio |
| Foreign-trained × reach interaction | 1.091 | 1.054 to 1.130 | <0.001 | Odds-ratio ratio |
| **Panel C. Cognitive-portfolio differences and associations with top-1% output** | | | | |
| Adjusted topic breadth, foreign minus U.S. | −0.0487 | −0.065 to −0.033 | <0.001 | SD difference |
| Cognitive span, foreign minus U.S. | −0.0426 | −0.056 to −0.029 | <0.001 | SD difference |
| Adjusted breadth association | 0.946 | 0.930 to 0.962 | <0.001 | Odds ratio |
| Cognitive-span association | 0.940 | 0.926 to 0.954 | <0.001 | Odds ratio |
| Breadth × span interaction | 1.037 | 1.018 to 1.056 | <0.001 | Odds ratio |
| Foreign-trained main effect | 1.067 | 1.024 to 1.112 | 0.002 | Adj. odds ratio |

*Notes.* Panel A models adjust for year, scientific domain, and institutional research intensity; N = 1,305,841 faculty-years. Panel B is a grouped-binomial model with author-clustered inference; N = 1,119,314 faculty-years and 7,630,784 publications. Panel C uses N = 853,822 faculty-years. Training origin is an adjustment variable, not part of the breadth × span interaction.

# 5. Robustness and heterogeneity analyses

## 5.1. Alternative temporal definitions of collaboration

The main network measure uses collaborations observed during the three years preceding each outcome year. We additional test alternative window sizes. In window-specific samples, the foreign-trained × reach interaction is 1.081 for a one-year window (95% CI, 1.047–1.116), 1.091 for the three-year main specification (95% CI, 1.054–1.130), and 1.099 for a five-year window (95% CI, 1.059–1.140). The corresponding estimates on the common 980,186-author-year sample are 1.081, 1.091, and 1.098. The main result is therefore robust across alternative temporal definition. Thus, across network windows, a one-SD increase in log network reach is associated with an approximately 8–10% larger multiplicative increase in the odds of top-1% output among foreign-trained than U.S.-trained faculty.

The stability is substantively useful because the windows represent different concepts. A one-year network emphasizes active collaborators and may be sensitive to short-term project cycles. A five-year network incorporates older ties and is more likely to capture durable professional reach. Similar interaction estimates across these specifications suggest that the result reflects a persistent feature of collaboration history.

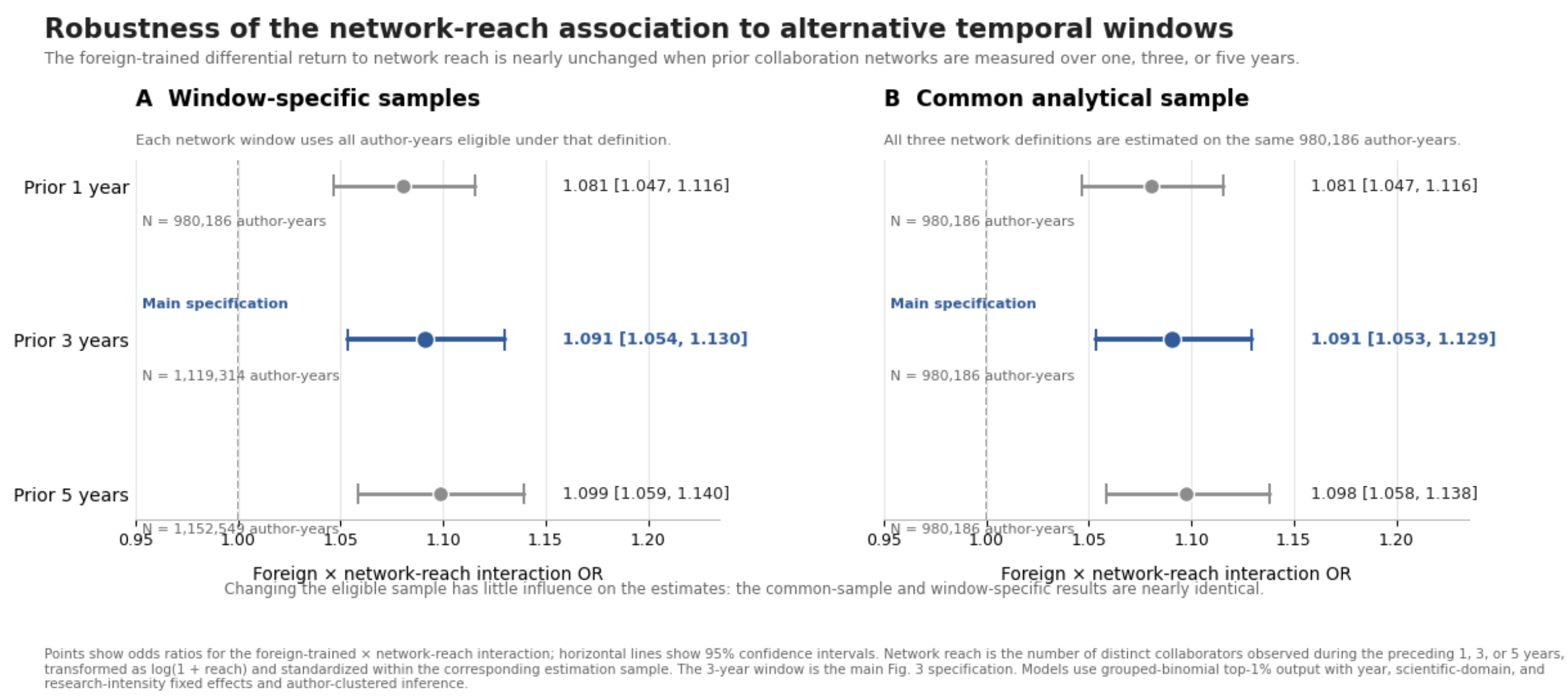

**Fig. 5.** Robustness to alternative temporal definitions of collaboration networks. (A) uses each one-, three-, and five-year network definition on its full eligible sample. (B) estimates all three definitions on the same 980,186 faculty-years. Points are foreign-trained × reach interaction odds ratios and horizontal lines are 95% confidence intervals. Each log-reach measure is standardized within its estimation sample; the three-year window is the main specification.

## 5.2. Scientific-domain heterogeneity

Heterogeneity analysis show that raw network position varies across fields (Fig. 6A). Medicine and Health and Natural Sciences have the largest prior networks for both foreign- and U.S.-trained groups, whereas Education and Business and Economics have much smaller networks. Foreign-trained faculty have higher mean reach in most domains, except for Education, Humanities, and Mathematics and Computing.

The adjusted interaction estimates also vary descriptively (Fig. 6B). They range from 0.967 in Education to 1.197 in Mathematics and Computing, though only two domains have estimates statistically distinguishable from 1, including Mathematics and Computing and in Medicine and Health, meaning that an increase in network reach is associated with a larger return on elite output for foreign-trained faculty than U.S.-trained faculty for faculty in these two domains. An 8-degree-of-freedom Wald test of the reach × origin × domain terms yields χ² = 11.98 and p = 0.152. We therefore find no strong evidence that the interaction systematically differs across the nine broad domains. Supplementary Table A3 reports the complete domain estimates.

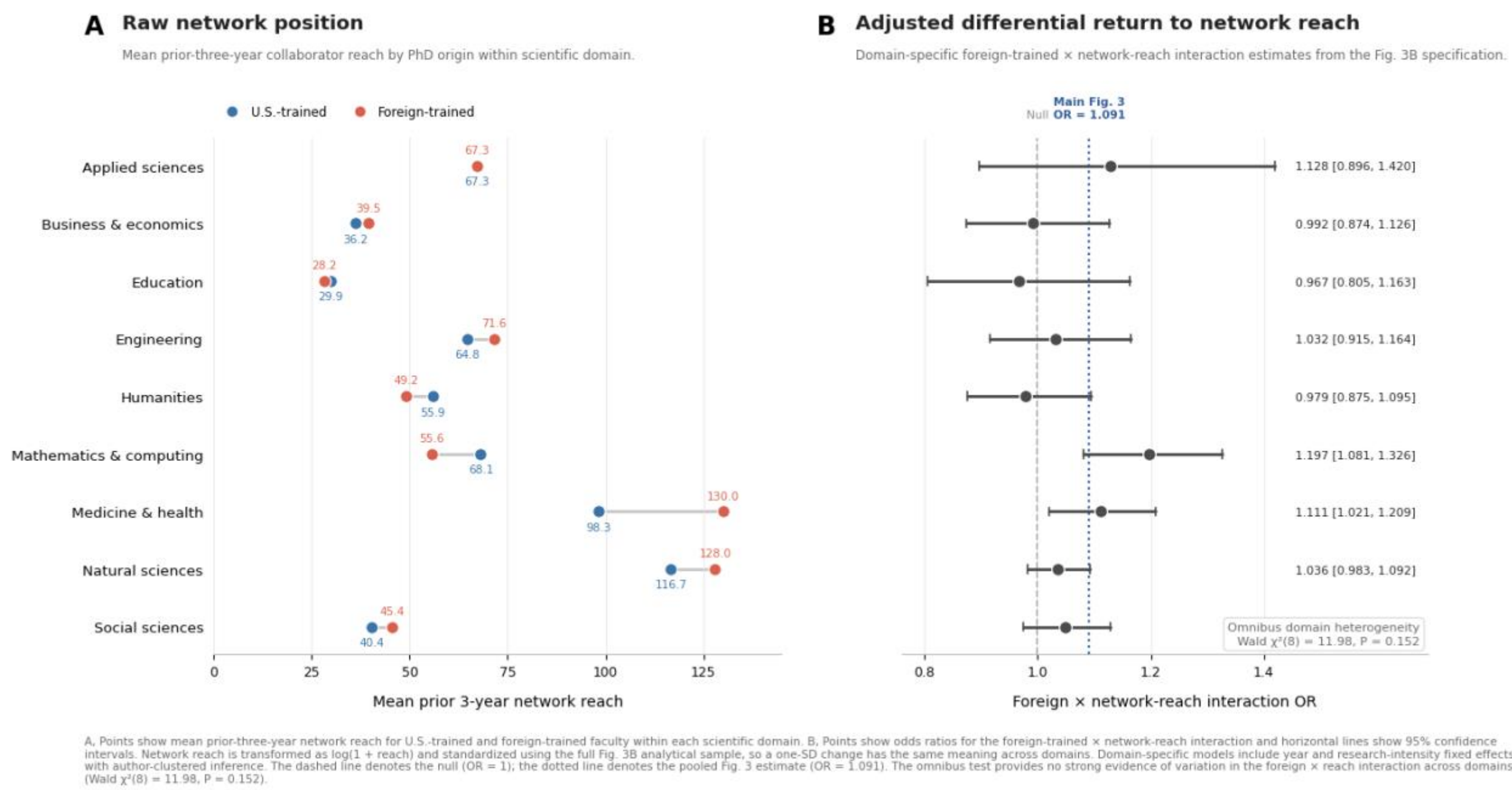

**Fig. 6.** Scientific-domain variation in network position and differential association with top-1% output. (A) Mean prior-three-year reach by training origin within nine scientific domains. (B) Domain-specific foreign-trained × reach interaction odds ratios with 95% confidence intervals, using the main-model reach scale. The dashed line denotes the null and the dotted line the pooled estimate. A joint Wald test provides no strong evidence of domain heterogeneity ($\chi^2(8) = 11.98$, p = 0.152).

## 5.3. Heterogeneity across foreign training systems

Country-specific results reveal more pronounced variation. Figure 7A shows large differences in mean prior reach among faculty trained in the 12 largest foreign terminal-degree systems. Italy, Germany, Japan, the United Kingdom, Switzerland, China, and a few others have means above the U.S.-trained benchmark, while France and Russia are below it.

The differential reach association also varies (Fig. 7B). Interaction odds ratios range from 0.845 for Switzerland to 1.284 for Israel. Estimates are above 1 and statistically significant for the United Kingdom, Canada, India, and Israel; other estimates are less precise, meaning that an increase in network reach is associated with a larger return on elite output for foreign-trained faculty than U.S.-trained faculty. Meanwhile, Switzerland has estimate less than 1, suggesting that for faculty-trained in Swiss institutions, a gain in network reach has a reduced return on elite output for foreign-trained faculty. A joint 11-degree-of-freedom test rejects equality of the 12 country-specific interactions ($\chi^2 = 27.16$, p = 0.004), and a 12-degree-of-freedom test rejects the null that all are jointly zero ($\chi^2 = 40.55$, $p = 5.83 \times 10^{-5}$).

The country analysis indicates that the network–impact association is not uniform across training origins. The estimates can reflect selective migration, field composition, institutional destinations, cohort histories, language, and the subset of each country's graduates who obtained U.S. faculty appointments. Supplementary Table A4 reports the complete country estimates.

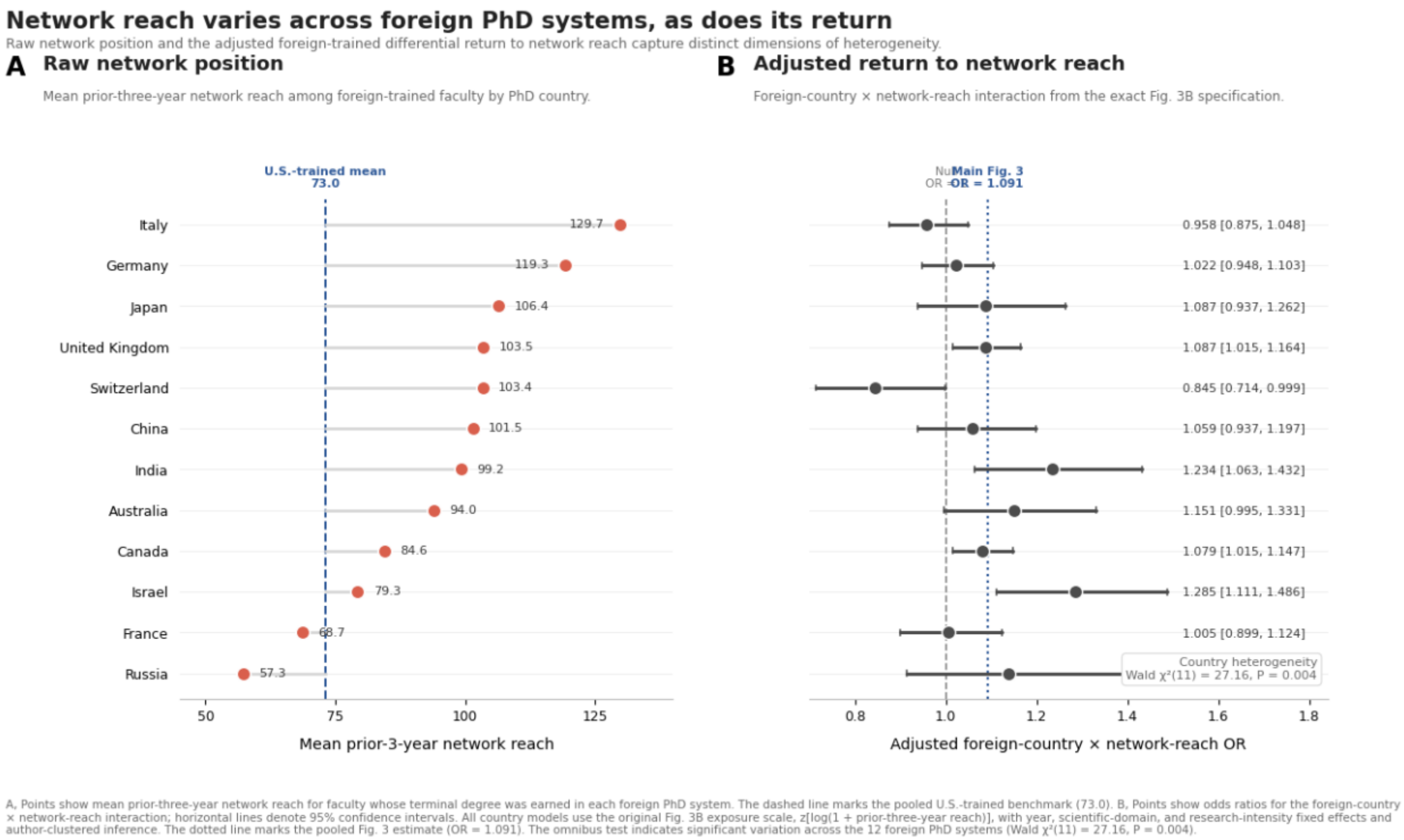

**Fig. 7.** Heterogeneity across major foreign terminal-degree systems. (A) Mean prior-three-year reach among faculty trained in the 12 largest foreign systems with at least 500 observed faculty; the dashed line is the U.S.-trained mean. (B) Country-specific interaction odds ratios relative to the common U.S.-trained reference, with 95% confidence intervals; the dotted line is the pooled foreign-trained estimate. A joint Wald test rejects equality of the 12 country-specific interactions ($\chi^2(11) = 27.16$, $p = 0.004$). Estimates describe associated heterogeneity and are not causal rankings.

## 5.4. Additional robustness checks

Three additional checks are reported in Supplementary Appendix (Supplementary Figs. A1–A3 and Table A2). First, the network interaction is positive for assistant, associate, and full professors, with estimates of 1.128, 1.027, and 1.093, respectively. The assistant and full professor estimates are statistically distinguishable from 1; the associate professor estimate is smaller and imprecise. Second, the estimate is 1.089 in highest-research institutions and 1.068 in other classified institutions. Third, replacing top-1% output with the broader top-10% outcome on the sample yields an interaction of 1.052 (95% CI, 1.030–1.074), compared with 1.091 for the top-1%. The attenuation is consistent with a stronger differential association at the extreme upper tail, and the results hold for the lesser upper tail. The tests show that the main network pattern is stable to alternative temporal measurement, a broader impact threshold, and major organizational or career-stage partitions.

# 6. Discussion

## 6.1. Mobility advantages are organizationally situated

Research on scientific mobility has established that internationally mobile scientists often publish more, receive greater recognition, and maintain more international collaborations (Cruz-Castro & Sanz-Menéndez, 2010; Jonkers & Cruz-Castro, 2013). More recent work has provided more contextualized interpretations by showing that mobility outcomes depend on career trajectories, organizational contexts, and the networks through which scientists access knowledge (Benz & Larivière, 2025; Chen et al., 2025). Capponi and Frenken (2025) similarly demonstrate that the consequences of academic training depend on the institutional and intellectual context in which training occurs. Our findings extend this contextual turn from mobility trajectories to destination-system structure. Foreign-trained faculty contribute disproportionately to highly cited U.S. research, but citation differences narrow sharply when faculty are compared within the same institution, domain, rank, and year. At the same time, differences in publication volume and mixed-training-origin collaboration remain. Thus, the foreign-training advantage contains two analytically distinct components: differential placement within the research system and differences that persist within comparable organizational settings.

This distinction modifies how mobility premiums should be interpreted. Organizational position is not merely a confounder separating foreign training from its “net” effect. Internationally trained researchers enter a highly stratified destination system, and their observed contribution partly reflects where that system places them. The meaningful objective is therefore not to detach mobility from context, but to situate it with comparable scientific capabilities to organizational environments. Accordingly, our study connects mobility research with organizational cumulative advantage and academic hierarchy (Clauset et al., 2015; Wapman et al., 2022; Way et al., 2019).

## 6.2. International mobility carries relational capital, not simply additional ties

The collaboration network results expand the literature on the benefit of having elevated scientific networks. Existing literature found migrant and return scientists possess larger international networks and international connections can constitute scientific social capital by providing access to foreign knowledge and collaborators (Franzoni et al., 2014; Scellato et al., 2015). Our results add a different dimension: foreign-trained faculty not only have greater prior collaborator reach and breadth, they also received higher return on elite output because of their wider collaboration compared with U.S.-trained faculty. This pattern is stable across alternative network windows and broadly similar across scientific domains. The result extends the literature from network possession to the relationship between network position and subsequent scientific performance.

That distinction matters for social capital. Our evidence suggests that the relationship between connectivity and scientific output depends partly on the trajectory through which that connectivity was accumulated. One possibility, consistent with Scellato et al. (2015), is that networks spanning training systems contain complementary or less redundant resources. The heterogeneity across terminal-degree countries reinforces this point. Our country results showed substantial variation in the network–output relationship across training systems. Such systems may differ in their connectivity to U.S. science, institutional pathways, and migration selectivity. International training should therefore be understood as a heterogeneous source of relational capital.

## 6.3. Mobility and knowledge recombination

A third explanation links mobility to knowledge recombination. International mobility exposes scientists to new research environments, collaborators, and ideas, potentially enabling exploration beyond established knowledge bases. A recent study has shown that mobility increased research quality alongside collaboration with new, higher-quality coauthors and exploration of new topics (Baruffaldi et al., 2020), which complements broader research linking novel knowledge combinations to scientific impact (Uzzi et al., 2013).

Our results place an important boundary on this explanation. Foreign-trained faculty appear cognitively broader in raw comparisons, but this difference disappears once the larger size of their prior publication portfolios is considered. Their distinctive collaboration reach therefore does not translate into correspondingly broader individual topic portfolios.

This separation of social reach from cognitive breadth suggests that access to heterogeneous knowledge may instead be organized through collaborators. In this sense, our findings complement the recombination argument, which suggests that one consequence of international training may reside in the social distribution of knowledge across relationships, rather than in greater cognitive diversification within individual portfolios (Wang, 2016).

### 6.4. International training as a feature of scientific capacity

These results connect literatures that have largely examined different parts of the same process. Mobility research has established that scientists carry knowledge and relationships across borders (Cruz-Castro & Sanz-Menéndez, 2010; Edler et al., 2011; Franzoni et al., 2014; Jonkers & Cruz-Castro, 2013; Scellato et al., 2015). Research on scientific careers shows that the consequences of mobility and training depend on trajectories and contexts (Baruffaldi & Landoni, 2012; Baruffaldi et al., 2020; Benz & Larivière, 2025). Network research, meanwhile, shows that access to international connections can affect scientific performance (Chen et al., 2025). Our study brings these arguments together at the level of the destination research system. Its principal contribution is to show that international training is associated with three separable features of scientific capacity: organizational structure, collaboration networks, and research portfolio.

### 6.5. Limitations and future research

Foreign-trained faculty employed at U.S. universities are selected through training, migration, hiring, and career progression, while network position is itself shaped by prior productivity, reputation, resources, and scientific specialization. The analyses identify organizational and relational patterns, not selection mechanisms of foreign-trained faculty into U.S. universities or causal effects of foreign training or collaborator reach. Terminal-degree location captures only one element of mobility history. Studies showed the importance of using complete career trajectories in measuring mobility effect (Benz & Larivière, 2025) and the stage at which international movement occurs is related to subsequent network structure (Scellato et al., 2015). Future research should therefore reconstruct sequences of doctoral training, postdoctoral mobility, institutional movement, and international collaboration. Last, coauthorship captures only publication ties and does not reveal other relationships such as mentorship or tacit knowledge transfer. These constraints provide boundaries of the interpretations of the obtained results.

## Data availability

AARC faculty data are licensed and cannot be redistributed by the authors. OpenAlex data are publicly available. Analytical SQL and Python code and nonproprietary derived aggregate outputs is available at https://github.com/erjiayan/foreign-trained-faculty.

## CRediT authorship contribution statement

Erjia Yan: Conceptualization, Methodology, Software, Formal analysis, Data curation, Visualization, Writing – original draft, Writing – review and editing. Chaoqun Ni: Conceptualization, Methodology, Supervision, Writing – original draft, Writing – review and editing. Xiang Zheng: Conceptualization, Methodology, Validation, Writing – original draft, Writing – review and editing. Weiye Gu: Methodology, Validation, Writing – original draft, Writing – review and editing.

## Declaration of competing interest

The authors declare no competing interest.

## Declaration of generative AI in the writing process

During the preparation of this work, the authors used ChatGPT (version 5.6) to assist with code development and language editing. The authors reviewed and edited all outputs and take full responsibility for the content of the publication.

# Supplementary Appendix

## A1. Analytical samples and linkage diagnostics

Table A1 reports the denominators used by the main and auxiliary analyses. It also records the confidence thresholds used for the AARC-to-OpenAlex entity resolution. Definitions and model equations are provided in the main text rather than repeated here.

**Table A1.** Analytical samples and linkage diagnostics

| Analysis | Analytical unit | Observations / strata | Faculty |
|---|---|---|---|
| System representation | Annual aggregates | 2011–2020 panels | 236,394 |
| Within-context comparison | Institution × domain × rank × year | 38,223 strata | — |
| Network position | Faculty-year | 1,305,841 | 213,119 |
| Network and top-1% output | Faculty-year | 1,119,314 | 200,348 |
| Cognitive portfolios | Faculty-year | 853,822 | 165,990 |
| Career-stage robustness | Faculty-year | 1,115,056 | 200,220 |
| Institutional robustness | Faculty-year | 1,113,110 | 199,250 |
| Country omnibus model | Faculty-year | 1,096,184 | 195,955 |

*Notes.* The author–work spine contains 12,109,441 linked faculty–publication observations for 209,959 U.S.-trained and 26,435 foreign-trained faculty. High-confidence links require score $S \geq 0.80$ and separation margin $\Delta \geq 0.05$; medium-confidence links require $0.50 \leq S < 0.80$ and $\Delta \geq 0.01$. Medium- or high-confidence coverage is approximately 81%.

## A2. Additional robustness checks

The following figures retain the three most informative checks not moved into the main manuscript: career-stage heterogeneity, institutional-environment heterogeneity, and robustness to a broader high-impact threshold. Table A2 places their focal estimates on a common scale.

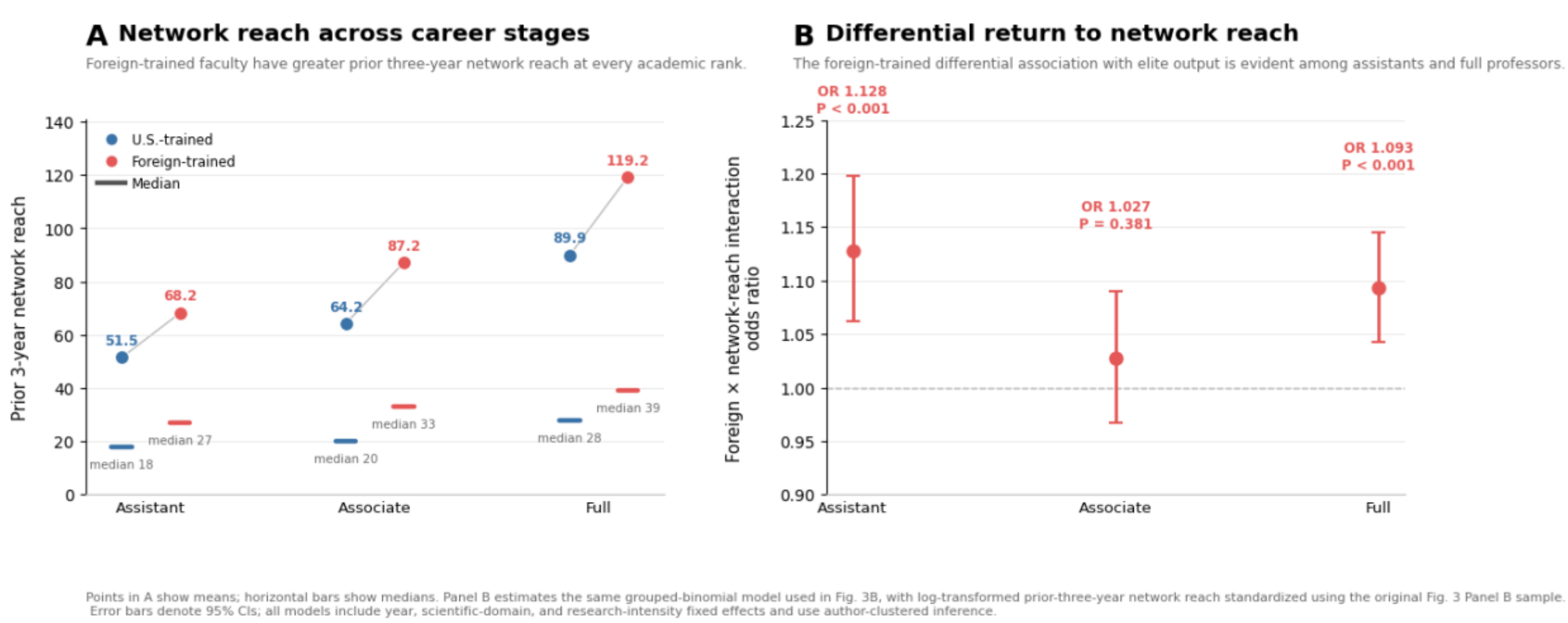

**Fig. A1.** Career-stage heterogeneity in network position and differential return. (A) Mean and median prior-three-year reach by training origin among assistant, associate, and full professors. (B) Foreign-trained × reach interaction odds ratios from the main grouped-binomial specification, with 95% confidence intervals. The rank-linked sample retains 99.62% of main-model faculty-years.

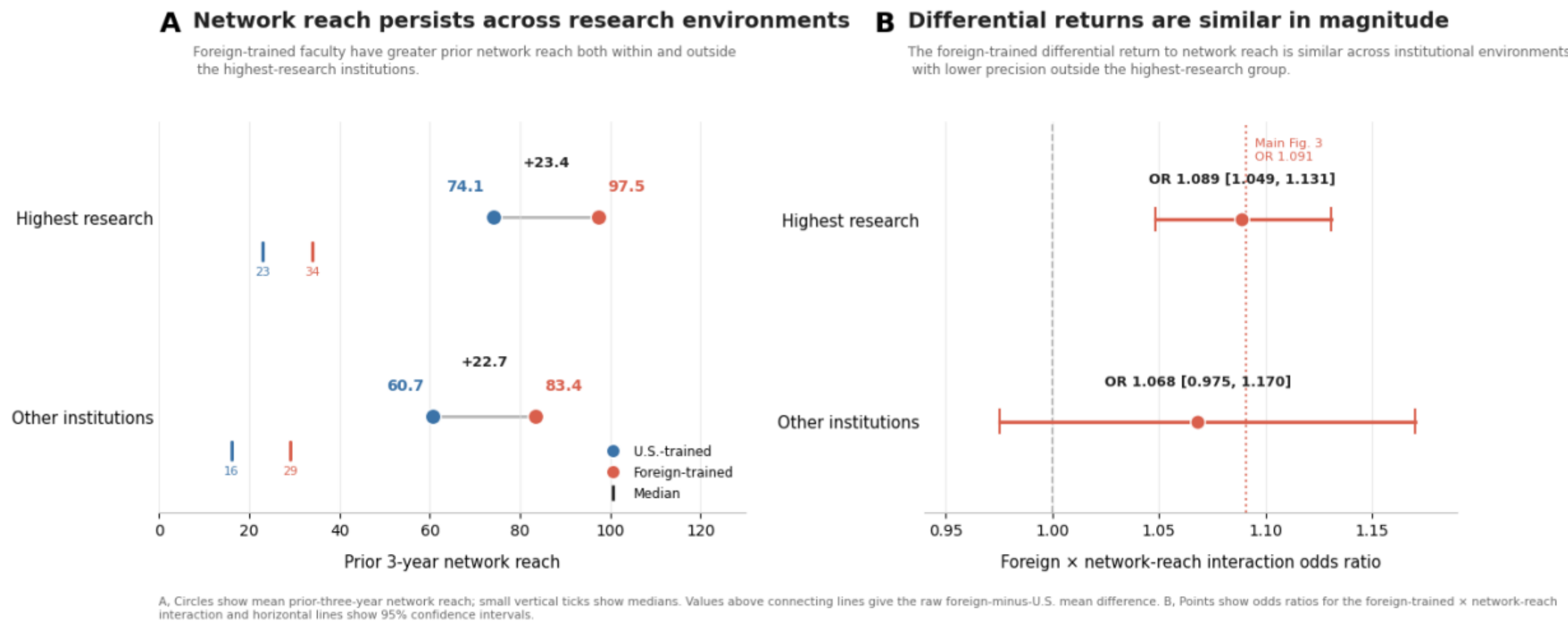

**Fig. A2.** Network position and differential return across institutional research environments. (A) Mean and median reach for U.S.-trained and foreign-trained faculty in highest-research and other classified institutions. (B) Foreign-trained × reach interaction odds ratios and 95% confidence intervals. The other-institution estimate is less precise because the stratum is substantially smaller.

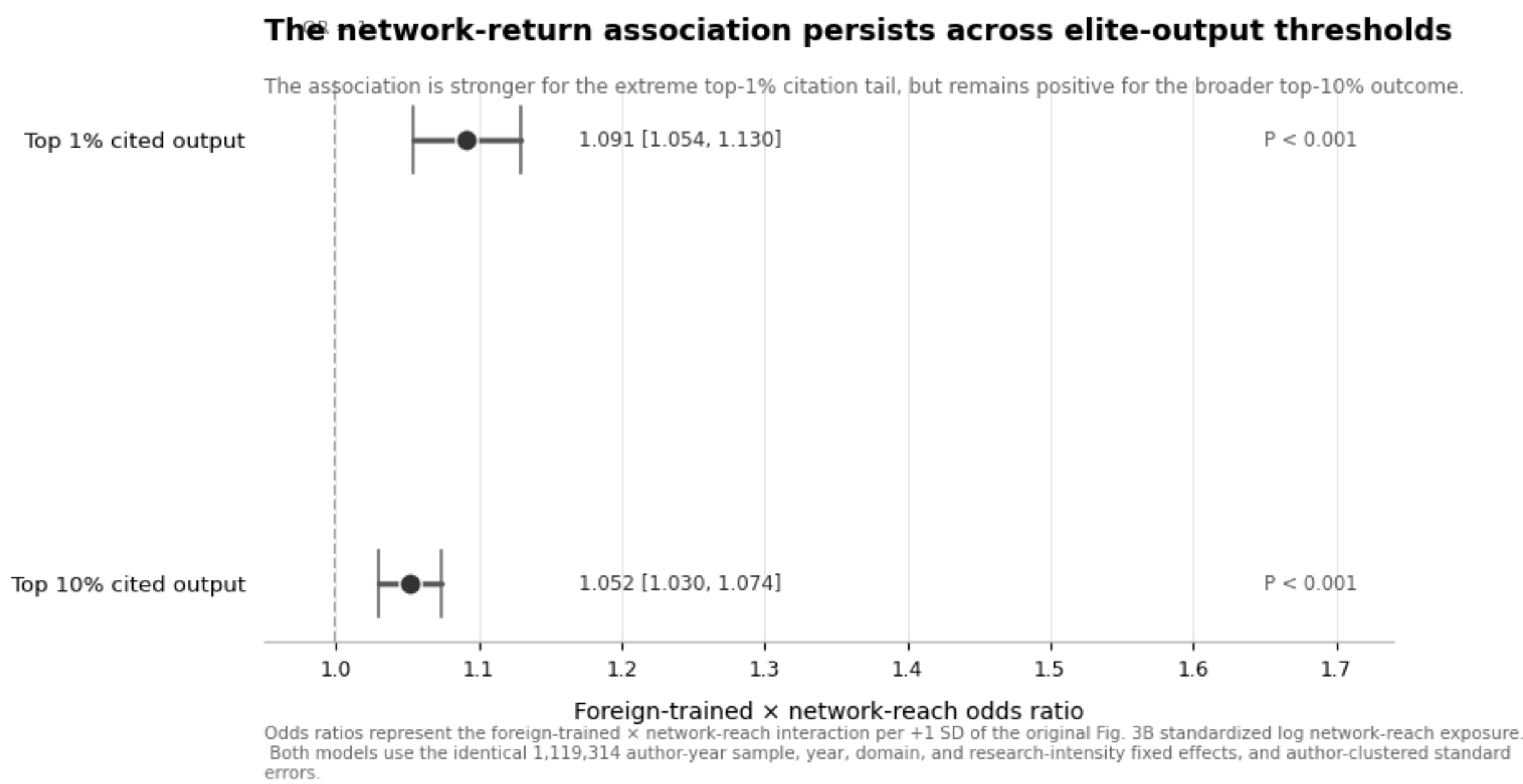

**Fig. A3.** Robustness to the elite-output threshold. Foreign-trained × reach interaction odds ratios are shown for top-1% and top-10% cited output using the identical 1,119,314 faculty-year sample, exposure scale, fixed effects, and clustered inference. The interaction is stronger at the top-1% threshold but remains positive for the broader top-10% outcome.

**Table A2.** Additional robustness and heterogeneity estimates

| Specification | Interaction OR | 95% confidence interval | p | Faculty-years |
|---|---|---|---|---|
| Main: 3-year reach, top-1% output | 1.091 | 1.054 to 1.130 | <0.001 | 1,119,314 |
| Assistant professors | 1.128 | 1.062 to 1.198 | <0.001 | 286,987 |
| Associate professors | 1.027 | 0.968 to 1.090 | 0.381 | 305,971 |
| Full professors | 1.093 | 1.043 to 1.145 | <0.001 | 522,098 |

| Specification | Interaction OR | 95% confidence interval | p | Faculty-years |
|---|---|---|---|---|
| Highest-research institutions | 1.089 | 1.049 to 1.131 | <0.001 | 991,308 |
| Other classified institutions | 1.068 | 0.975 to 1.170 | 0.155 | 121,802 |
| Top-10% output, identical sample | 1.052 | 1.030 to 1.074 | <0.001 | 1,119,314 |

*Notes*. All estimates are foreign-trained × standardized log network-reach interaction odds ratios from the grouped-binomial specification. Rank models use the main-sample reach scale. The top-10% model holds the author-year sample, exposure scale, fixed effects, and clustered inference constant.

## A3. Complete field and training-system estimates

Figures 6 and 7 in the main manuscript present the substantive patterns. Tables A3 and A4 provide the complete numerical estimates used in those displays. The joint tests, rather than isolated p-value comparisons, govern the interpretation of heterogeneity.

**Table A3.** Scientific-domain heterogeneity in network position and differential return

| Domain | U.S. mean reach | Foreign mean reach | Faculty-years | Foreign faculty-years | Interaction OR (95% CI) | p |
|---|---|---|---|---|---|---|
| Applied Sciences | 67.3 | 67.3 | 39,462 | 3,999 | 1.128 (0.896–1.420) | 0.306 |
| Business and Economics | 36.2 | 39.5 | 103,168 | 5,871 | 0.992 (0.874–1.126) | 0.902 |
| Education | 29.9 | 28.2 | 47,700 | 1,252 | 0.967 (0.805–1.163) | 0.724 |
| Engineering | 64.8 | 71.6 | 160,823 | 18,169 | 1.032 (0.915–1.164) | 0.610 |
| Humanities | 55.9 | 49.2 | 96,994 | 8,578 | 0.979 (0.875–1.095) | 0.707 |
| Mathematics and Computing | 68.1 | 55.6 | 87,069 | 12,947 | 1.197 (1.081–1.326) | <0.001 |
| Medicine and Health | 98.3 | 130.0 | 195,708 | 25,126 | 1.111 (1.021–1.209) | 0.015 |
| Natural Sciences | 116.7 | 128.0 | 248,291 | 45,630 | 1.036 (0.983–1.092) | 0.188 |
| Social Sciences | 40.4 | 45.4 | 140,099 | 6,862 | 1.049 (0.975–1.128) | 0.199 |

*Notes.* Domain-specific models use the main-model reach scale and adjust for year and institutional research intensity. The omnibus test of reach × origin × domain terms is Wald $\chi^2(8) = 11.98$, $p = 0.152$.

**Table A4.** Network position and differential return by foreign terminal-degree system

| Terminal-degree country | Foreign faculty-years | Foreign faculty | Mean reach | Top-1% rate | Interaction OR (95% CI) | p |
|---|---|---|---|---|---|---|
| United Kingdom | 25,009 | 4,440 | 103.5 | 5.389% | 1.086 (1.014–1.163) | 0.018 |
| Canada | 21,657 | 3,962 | 84.6 | 4.843% | 1.079 (1.015–1.148) | 0.014 |
| Germany | 11,484 | 1,962 | 119.3 | 4.657% | 1.022 (0.948–1.102) | 0.569 |
| China | 7,977 | 1,564 | 101.5 | 4.260% | 1.057 (0.936–1.195) | 0.371 |
| India | 7,834 | 1,424 | 99.2 | 3.367% | 1.234 (1.064–1.432) | 0.006 |
| Russia | 6,035 | 931 | 57.3 | 3.578% | 1.138 (0.916–1.414) | 0.243 |
| France | 4,830 | 890 | 68.7 | 4.707% | 1.005 (0.899–1.123) | 0.932 |
| Japan | 4,436 | 766 | 106.4 | 4.227% | 1.085 (0.935–1.260) | 0.283 |
| Italy | 4,353 | 814 | 129.7 | 5.754% | 0.959 (0.876–1.049) | 0.359 |
| Australia | 4,274 | 786 | 94.0 | 5.981% | 1.149 (0.994–1.328) | 0.061 |
| Israel | 4,125 | 661 | 79.3 | 5.869% | 1.284 (1.110–1.486) | <0.001 |
| Switzerland | 3,290 | 594 | 103.4 | 4.694% | 0.845 (0.714–0.999) | 0.048 |

*Notes.* Each country-specific coefficient compares the focal foreign system with the common U.S.-trained reference. Equality of the 12 interactions: Wald $\chi^2(11) = 27.16$, $p = 0.004$. All 12 interactions jointly zero: Wald $\chi^2(12) = 40.55$, $p = 5.83 \times 10^{-5}$. Estimates are associational and should not be interpreted as national rankings.